\documentclass[aps,prl,showpacs,twocolumn,groupedaddress]{revtex4}

\usepackage{graphicx}
\usepackage{dcolumn}
\usepackage{bm}

\begin{document}

\preprint{APS/123-QED}

\title{Magnetic Properties of the Spin-1/2 Deformed Kagome Antiferromagnet Edwardsite}

%

\author{Hajime Ishikawa, Yoshihiko Okamoto, and Zenji Hiroi}
\affiliation{
Institute for Solid State Physics, University of Tokyo, Kashiwa 277-8581, Japan
}

\date{\today}

\begin{abstract}
We prepared a powder sample of edwardsite Cd$_2$Cu$_3$(SO$_4$)$_2$(OH)$_6$$\cdot$4H$_2$O, which is a new candidate compound for the spin-1/2 kagome antiferromagnet, and studied its magnetic properties by magnetic susceptibility and heat capacity measurements. Edwardsite has a deformed kagome lattice with an average antiferromagnetic interaction of 51 K between nearby spins and shows an antiferromagnetic order accompanied by a small ferromagnetic moment below 4.3 K. The weak ferromagnetism is likely due to spin canting caused by sizable Dzyaloshinsky-Moriya interactions, which may stabilize the long-range magnetic order instead of a spin-liquid state expected for the kagome antiferromagnet.
\end{abstract}

\pacs{Valid PACS appear here}
\maketitle

The spin-1/2 antiferromagnet on a kagome lattice consisting of a two-dimensional network of corner-sharing triangles is one of the most intensively studied frustrated spin systems. It is theoretically expected that an exotic ground state such as a gapped or gapless spin-liquid state or a valence-bond-solid state will be stable there owing to the combination of geometrical frustration and quantum fluctuation~\cite{theory}. In a magnetic field, moreover, intriguing phenomena such as a 1/3 magnetization plateau and a magnetization jump from 7/9 to the saturation are predicted to appear~\cite{Plateau}. Two Cu minerals, herbertsmithite~\cite{Herb1,Herb2} and vesignieite~\cite{Vesi1}, have been investigated as model compounds. The ground state of the former is believed to be a spin-liquid state, while that of the latter is the 120$^{\circ}$ ordered state with a \textbf{q} = 0 propagation vector, which may be stabilized by a relatively strong Dzyaloshinsky-Moriya (DM) interaction~\cite{Vesi2,Vesi3,Vesi4}. 

There are more Cu compounds in nature with deformed kagome lattices, in which the effect of the deformation of the kagome lattice has been examined in order to obtain insight into the true ground state of the kagome antiferromagnet (KAFM). A typical example is volborthite Cu$_3$V$_2$O$_7$(OH)$_2$$\cdot$2H$_2$O. It exhibits a phase transition to a peculiar magnetic phase with an extremely slow spin fluctuation at $\sim$1 K~\cite{Vol3,Vol4}. It also shows multiple phase transitions in a magnetic field that appear as steps in the magnetization curve~\cite{Vol2}. The result shows that the $H$-$T$ phase diagram of volborthite is complex, exemplifying the rich physics arising from frustration in a deformed kagome lattice. KCu$_3$As$_2$O$_7$(OH)$_3$ has a more deformed kagome lattice that includes both ferromagnetic and antiferromagnetic interactions. It shows an unusual antiferromagnetic order accompanied by a large entropy release well below the ordering temperature~\cite{KCuAs}. On the other hand, in Rb$_2$Cu$_3$SnF$_{12}$, a pinwheel-shaped deformation of Cu$^{2+}$ dodecamers results in a ``pinwheel'' valence-bond-solid state~\cite{Matan}. Thus, the effects of deformations on the kagome lattice are diverse, so that systematic studies of various kagome compounds are necessary to elucidate the key physics of the spin-1/2 KAFM.

\begin{figure}
\begin{center}
\includegraphics[width=8.5cm]{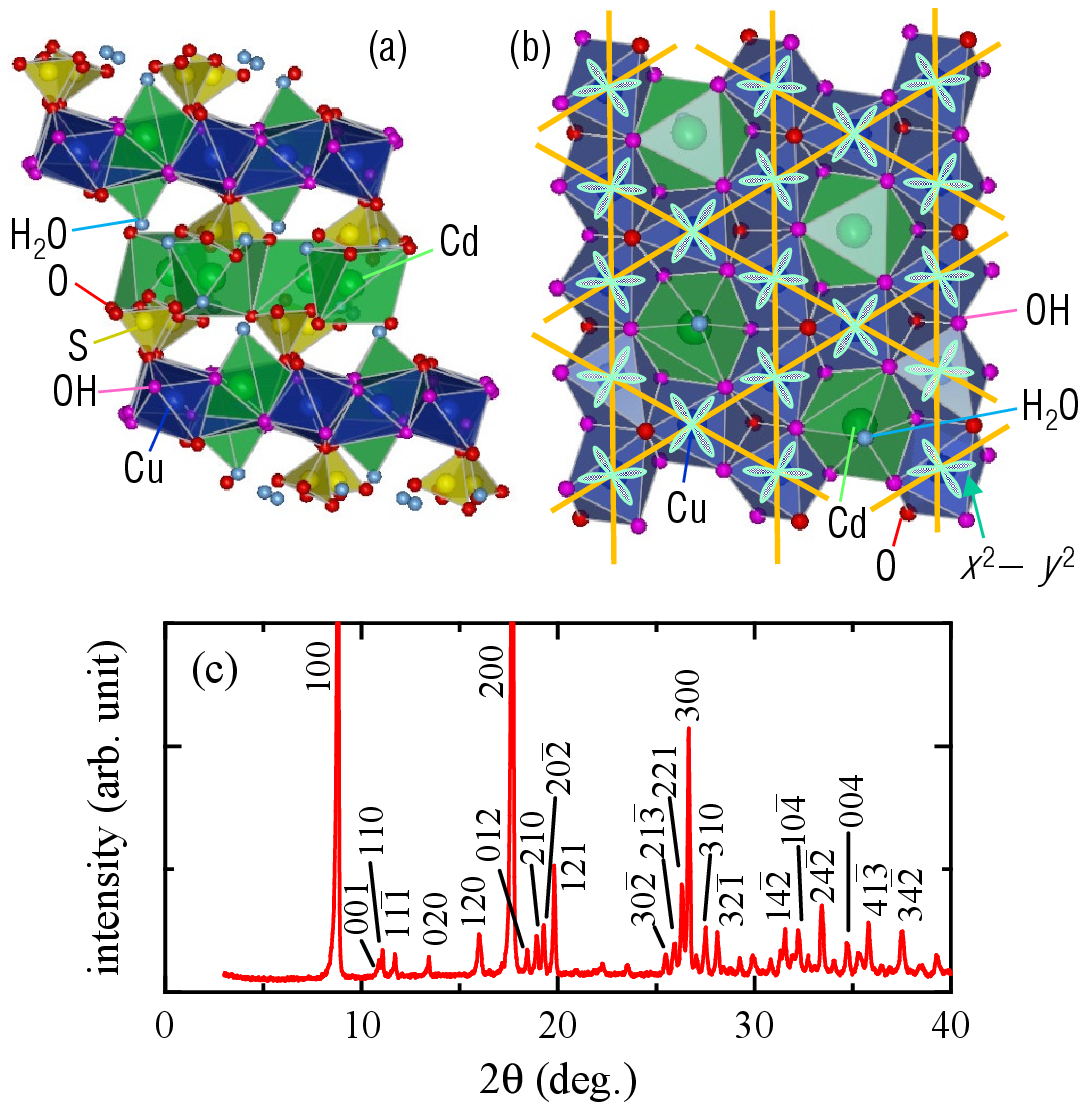}
\caption{Crystal structure of edwardsite viewed along the $b$ axis (a) and perpendicular to the $bc$ plane (b). The $x^2 - y^2$ orbitals occupied by an unpaired electron in each Cu$^{2+}$ ion are shown in (b). (c) XRD pattern of a powder sample of edwardsite taken at room temperature. Peak indices are given for a monoclinic unit cell with the lattice constants $a$ = 10.880(3) \r{A}, $b$ = 13.175(5) \r{A}, $c$ = 11.208(4) \r{A}, and $\beta$ = 112.95(2)$^{\circ}$.}
\label{F1}
\end{center}
\end{figure}

In this letter, we report the magnetic properties of another Cu mineral edwardsite, which has not been focused on thus far as a candidate compound for the spin-1/2 antiferromagnet. Edwardsite, Cd$_2$Cu$_3$(SO$_4$)$_2$(OH)$_6$$\cdot$4H$_2$O, is a natural mineral recently identified by Elliott \textit{et al}~\cite{Elliott}. It crystallizes in a monoclinic structure with the space group of $P$2$_1$/$c$; the lattice constants are $a$ = 10.863(2) \r{A}, $b$ = 13.129(3) \r{A}, $c$ = 11.169(2) \r{A}, and $\beta$ = 113.04(3)$^{\circ}$. 
The number of formula units per unit cell is $Z$ = 4. As shown in Fig. 1, edge sharing CuO$_4$(OH)$_2$ octahedra and Cd(OH)$_6$(H$_2$O) capped octahedra form a slab, in which Cu$^{2+}$ ions are aligned in a kagome geometry and Cd$^{2+}$ ions are located at the center of the hexagon of the kagome lattice. The kagome layers are well separated from each other along the stacking direction by a thick nonmagnetic block layer consisting of CdO$_3$(H$_2$O)$_3$ octahedra sandwiched by a couple of layers of SO$_4$ tetrahedra. The distance between kagome layers is 10.0 \r{A} in edwardsite, which is significantly larger than 4.7 \r{A} in herbertsmithite, 7.2 \r{A} in volborthite, and 6.9 \r{A} in vesignieite. These suggest that the interplanar coupling is relatively small in edwardsite. 

Another important structural feature of edwardsite is the unique deformation pattern of the kagome lattice. Cu$^{2+}$ ions occupy four crystallographic sites, and the Cu-Cu distance in the kagome lattice varies within 5\%~\cite{Elliott}. However, one expects that the associated modulation in magnetic interactions is rather small, because all the Cu$^{2+}$ spins are accommodated in $x^2 - y^2$ orbitals arranged so as to approximately maintain three-fold rotation axes in the kagome lattice, as shown in Fig. 1(b); the orbital selection is uniquely determined by the shape of the coordination octahedra of Cu$^{2+}$ ions. This orbital arrangement in edwardsite is the same as that in herbertsmithite but different from those in volborthite and KCu$_3$As$_2$O$_7$(OH)$_3$; the $x^2 - y^2$ and $3z^2 - r^2$ orbitals coexist in the latter two compounds~\cite{Vesi1,KCuAs}. Considering these structural features, we expect that edwardsite can be a good model compound. 
However, natural edwardsite may be unsuitable for studying the magnetism of a kagome lattice, because it contains considerable numbers of Zn and Fe atoms, which disorder the Cu kagome lattice~\cite{Elliott}. We prepared a pure sample of Cd$_2$Cu$_3$(SO$_4$)$_2$(OH)$_6$$\cdot$4H$_2$O and studied its physical properties. It is found that edwardsite has an average antiferromagnetic interaction of 51 K and exhibits an antiferromagnetic order accompanied by a weak ferromagnetic moment below 4.3 K. 

\begin{figure}
\begin{center}
\includegraphics[width=8cm]{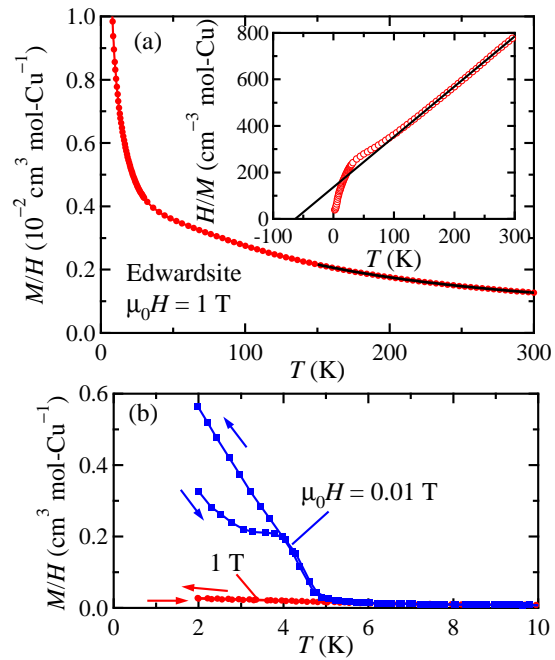}
\caption{(a) Temperature dependence of $M$/$H$ at a magnetic field of 1 T for a powder sample of edwardsite. Diamagnetic contributions from core electrons have already been subtracted from the data. The solid curve between 150 and 300 K represents a fit to a kagome lattice model obtained by high-temperature series expansion~\cite{HTSE}. The inset shows inverse susceptibility, where the solid line represents a Curie-Weiss fit. (b) Temperature dependence of zero-field-cooled and field-cooled $M$/$H$ values measured at two magnetic fields of 0.01 and 1 T. }
\label{F2}
\end{center}
\end{figure}

We prepared a powder sample of edwardsite by chemical reaction in a solution. 0.1 g of Cu(OH)$_2$, 1.5 g of 3CdSO$_4$$\cdot$8H$_2$O, and 0.2 g of (NH$_4$)$_2$SO$_4$ were put in a Pyrex beaker with 15 ml of water. The degrees of purity of these starting materials are at least 90, 99.9, and 99.5\%, respectively. The mixture was stirred and then kept at 65 $^{\circ}$C for 12 h. To improve its crystallinity and increase its grain size, the obtained powder was heated in CdSO$_4$ aqueous solution in a sealed stainless-steel vessel at 150 $^{\circ}$C for 12 h. A sky-blue powder was obtained after rinsing with water several times and drying at room temperature. Sample characterization was performed by powder X-ray diffraction (XRD) analysis with Cu K$\alpha$ radiation at room temperature, employing a RINT-2000 diffractometer (Rigaku). All diffraction peaks observed in the powder XRD pattern shown in Fig. 1(c) can be indexed on the basis of a monoclinic structure of space group $P$2$_1$/$c$ with a lattice consisting of $a$ = 10.880(3) \r{A}, $b$ = 13.175(5) \r{A}, $c$ = 11.208(4) \r{A}, and $\beta$ = 112.95(2)$^{\circ}$, which are close to those reported for a natural mineral by Elliott \textit{et al}~\cite{Elliott}. The peak width is slightly less than that of a high-quality sample of vesignieite~\cite{Vesi3}, indicating a relatively good crystallinity of the sample. Magnetic susceptibility and heat capacity were measured using a Magnetic Property Measurement System and a Physical Property Measurement System, respectively (both Quantum Design).

\begin{figure}
\begin{center}
\includegraphics[width=8cm]{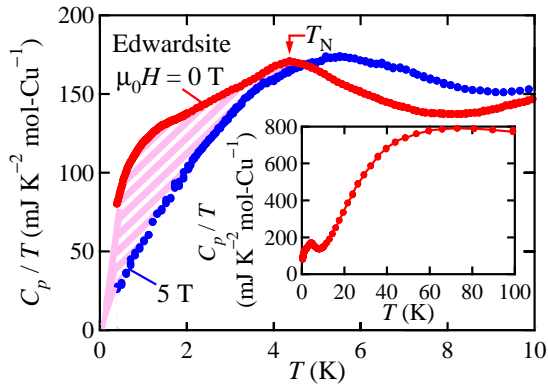}
\caption{Temperature dependence of heat capacity divided by temperature measured at magnetic fields of 0 and 5 T for a powder sample of edwardsite. $T_{\mathrm{N}}$ at $\mu_0H$ = 0 T is marked by an arrow. The inset shows the 0 T data up to 100 K. }
\label{F3}
\end{center}
\end{figure}

Figure 2(a) shows the temperature dependence of magnetic susceptibility measured on a powder sample of edwardsite. The diamagnetic contributions of core electrons of $\chi_{\mathrm{dia}}$ = $-$9.23 $\times$ 10$^{-5}$ cm$^{3}$ mol-Cu$^{-1}$ have already been subtracted from the data~\cite{LB}. The $H$/$M$ shown in the inset exhibits a linear dependence above 150 K, indicating that the Curie-Weiss law is obeyed. A fit to the equation $H$/$M$ = ($T - \theta_{\mathrm{W}}$)/$C$, where $C$ and $\theta_{\mathrm{W}}$ are the Curie constant and Weiss temperature, respectively, yields $C$ = 0.4661(5) cm$^{3}$ K mol-Cu$^{-1}$ and $\theta_{\mathrm{W}}$ = $-$66.2(3) K. The $C$ corresponds to an effective moment of $\mu_{\mathrm{eff}}$ = 1.93 $\mu_{\mathrm{B}}$ per Cu atom, while the $\theta_{\mathrm{W}}$ gives an average antiferromagnetic interaction of $J$/$k_{\mathrm{B}}$ = 66.2 K between nearby spins in the kagome lattice in the mean-field approximation: $\theta_{\mathrm{W}}$ = $-zJS(S + 1)$/3$k_{\mathrm{B}}$ with $z$ = 4. To obtain a more reliable $J$, we fit the $M$/$H$ data between 150 and 300 K to a calculation for the spin-1/2 KAFM using the high-temperature-series expansion, as shown in Fig. 2(a)~\cite{HTSE}. This fit gives $J$/$k_{\mathrm{B}}$ = 51.1(1) K and a Lande g-factor of $g$ = 2.195(1), which is close to that for vesignieite ($g$ = 2.14) determined by ESR measurement~\cite{ESR}. The magnitude of $J$ is smaller than those for herbertsmithite ($J$/$k_{\mathrm{B}}$ $\sim$ 170 K) and volborthite ($J$/$k_{\mathrm{B}}$ = 84 K), but close to that for vesignieite ($J$/$k_{\mathrm{B}}$ = 53 K)~\cite{Herb4,Vol1,Vesi1}. 

\begin{figure}
\begin{center}
\includegraphics[width=8cm]{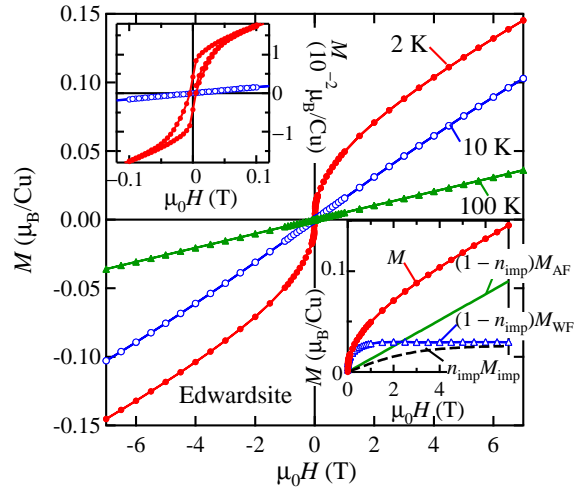}
\caption{Magnetization curves measured at 2, 10, and 100 K for a powder sample of edwardsite. The upper-left inset is an enlarged plot of the low-magnetic-field part of the 2 and 10 K curves. The lower-right inset shows $M$ measured at 2 K, which are decomposed into three components as described in the text: (1 $-$ $n_{\mathrm{imp}}$)$M_{\mathrm{AF}}$, (1 $-$ $n_{\mathrm{imp}}$)$M_{\mathrm{WF}}$, and $n_{\mathrm{imp}}$$M_{\mathrm{imp}}$.}
\label{F4}
\end{center}
\end{figure}

The $M$/$H$ values below 10 K at magnetic fields of 0.01 and 1 T are shown in Fig. 2(b). The 0.01 T data rapidly increase with decreasing temperature below 5 K, followed by a thermal hysteresis between the zero-field-cooled and field-cooled data below 4 K. This increase is completely suppressed at a magnetic field of 1 T, indicating that a weak ferromagnetic order takes place. In fact, an isothermal hysteresis, which is characteristic of a ferromagnet, is not observed in the $M$-$H$ curve measured at 10 K but clearly observed at 2 K, as shown in Fig. 4. Thus, the observed thermal hysteresis in Fig. 2(b) should not be due to a spin glass transition but related to the formation of ferromagnetic domains: ferromagnetic domains are generated below $T_{\mathrm{N}}$ and frozen so as to reduce the net moment when cooled at zero magnetic field, while they are forced to align when cooled in a magnetic field. The heat capacity $C_p$ data shown in Fig. 3 is also supportive of the presence of a bulk magnetic transition at this temperature: $C_p$/$T$ measured at zero magnetic field gradually increases with decreasing temperature below 8 K and shows a broad but distinct peak at $T_{\mathrm{N}}$ = 4.3 K. The broad nature of the transition might be due to the poor homogeneity of our sample. Since magnetic interactions in edwardsite are definitely antiferromagnetic, this magnetic phase transition must be due to a canted antiferromagnetic order, where Cu$^{2+}$ spins are slightly tilted so as to give an uncompensated magnetic moment. 

A magnetic response from a quantum spin compound is always contaminated by the presence of impurity spins. The amount of impurity spins can be a simple measure of sample quality. Moreover, since impurity spins give a magnetic susceptibility that increases divergently toward $T$ = 0, it is crucial to extract this extrinsic contribution in order to obtain intrinsic magnetic susceptibility at low temperatures. The previously reported values are 7.7-11\% for herbertsmithite~\cite{Herb5,Herb6}, 7\% for early sample of vesignieite~\cite{Vesi1}, and 0.07\% for volborthite~\cite{Vol2}. We estimate the amount of impurity spins in our edwardsite sample by analyzing the $M$-$H$ curve shown in Fig. 4. The $M$-$H$ curve at 2 K should consist of three components: an antiferromagnetic component $M_{\mathrm{AF}}$, a weak ferromagnetic one $M_{\mathrm{WF}}$ that must saturate above a low magnetic field, and $M_{\mathrm{imp}}$ from impurity spins. We fit the 2 K curve between 2 and 7 T to the equation $M$ = (1 $-$ $n_{\mathrm{imp}}$)($M_{\mathrm{AF}} + M_{\mathrm{WF}}$) + $n_{\mathrm{imp}}$$M_{\mathrm{imp}}$, where $n_{\mathrm{imp}}$ is the fraction of impurity spins, $M_{\mathrm{AF}}$ = $\chi_{\mathrm{AF}}H$, and $M_{\mathrm{imp}}$ = $N_{\mathrm{A}}gS\mu_{\mathrm{B}} B_{\mathrm{s}}(gS\mu_{\mathrm{B}}H$/$k_{\mathrm{B}}T$), assuming completely free spins whose $M$-$H$ curve is approximated by the Brillouin function $B_{\mathrm{s}}$. $g$ is fixed to be 2.195 as obtained from the $M$/$H$ data. Provided that $M_{\mathrm{WF}}$ is constant above 2 T, the fit is almost perfect and yields $n_{\mathrm{imp}}$ = 0.0261(4), $M_{\mathrm{WF}}$ = 0.0289(2) $\mu_{\mathrm{B}}$, and $\chi_{\mathrm{AF}}$ = 0.0125(4) $\mu_{\mathrm{B}}$ T$^{-1}$. Thus, our edwardsite sample contains 2.6\% impurity spins, less than those in herbertsmithite and vesignieite, but much greater than those in volborthite. 
\begin{figure}
\begin{center}
\includegraphics[width=8cm]{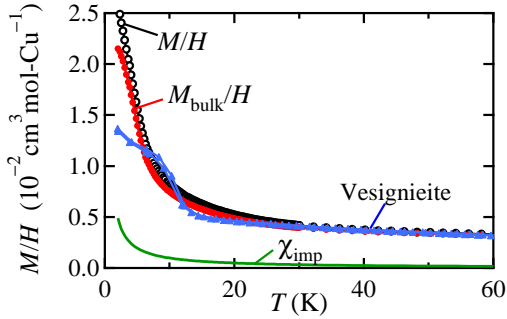}
\caption{Temperature dependence of the intrinsic magnetic susceptibility $M_{\mathrm{bulk}}$/$H$ of edwardsite measured at a magnetic field of 1 T, which is obtained by subtracting $\chi_{\mathrm{imp}}$ from $M$/$H$. The magnetic susceptibility of a polycrystalline sample of vesignieite is also shown for comparison, which shows an antiferromagnetic order below 9 K~\cite{Vesi3}. }
\label{F5}
\end{center}
\end{figure}

The presence of impurity spins should also manifest itself in heat capacity. As shown in Fig. 3, a shoulder at approximately 1 K in the $C_p$/$T$ measured at $\mu_0H$ = 0 T disappears at 5 T, indicative of an entropy shift to higher temperatures. The shifted entropy is roughly estimated to be 140 mJ mol-Cu$^{-1}$ K$^{-1}$ by subtracting the 5 T data from the zero field data (hatched area in Fig. 3). This entropy corresponds to 2.4\% of the total spin entropy of 1 mol of $S$ = 1/2 spins, i.e., $R$ln2 = 5.76 J mol$^{-1}$ K$^{-1}$, and is nearly equal to 2.6\% from the $M$-$H$ curve. The origin of the impurity spins in edwardsite is not clear at present, but they may be attributed to the lattice defects and/or the surface of small particles. Further optimization of sample preparation conditions is required to reduce impurity spins and attain higher-quality samples. 

Figure 5 shows the temperature dependence of the intrinsic magnetic susceptibility $M_{\mathrm{bulk}}$/$H$ obtained by subtracting the impurity spin contribution $\chi_{\mathrm{imp}}$ from $M$/$H$: $M_{\mathrm{bulk}}$/$H$ = ($M$/$H$ $-$ $\chi_{\mathrm{imp}}$)/(1 $-$ $n_{\mathrm{imp}}$) = ($M$/$H$ $-$ $n_{\mathrm{imp}}N_{\mathrm{A}}g^2\mu_{\mathrm{B}}^2$/4$k_{\mathrm{B}}$)/(1 $-$ $n_{\mathrm{imp}}$). $M_{\mathrm{bulk}}$/$H$ strongly increases with decreasing temperature below 30 K, reflecting the development of weak ferromagnetic correlations.

We compare the magnetic properties of edwardsite and vesignieite with similar $J$ values; that is, $J$/$k_{\mathrm{B}}$ = 51 and 53 K, respectively. Vesignieite exhibits a canted-antiferromagnetic order at $T_{\mathrm{N}}$ = 9 K. The magnetic structure is basically a 120$^{\circ}$ order with a \textbf{q} = 0 propagation vector, as evidenced by $^{51}$V-NMR measurement~\cite{Vesi3}. The accompanying weak-ferromagnetic moment is $M_{\mathrm{WF}}$ $\sim$ 0.015 $\mu_{\mathrm{B}}$ from the $M$-$H$ curve measured at 2 K~\cite{Vesi4}. As shown in Fig. 5, the magnetic susceptibilities of edwardsite and vesignieite are similar: both strongly increase above $T_{\mathrm{N}}$'s and tend to saturate toward $T$ = 0. Moreover, the weak-ferromagnetic moment of $M_{\mathrm{WF}}$ = 0.0289(2) $\mu_{\mathrm{B}}$ in edwardsite is quite small and close to that in vesignieite. Therefore, it is quite reasonable to assume that the magnetic order in edwardsite is a canted-antiferromagnetic order, the same as in vesignieite. 

\begin{figure}
\begin{center}
\includegraphics[width=8.5cm]{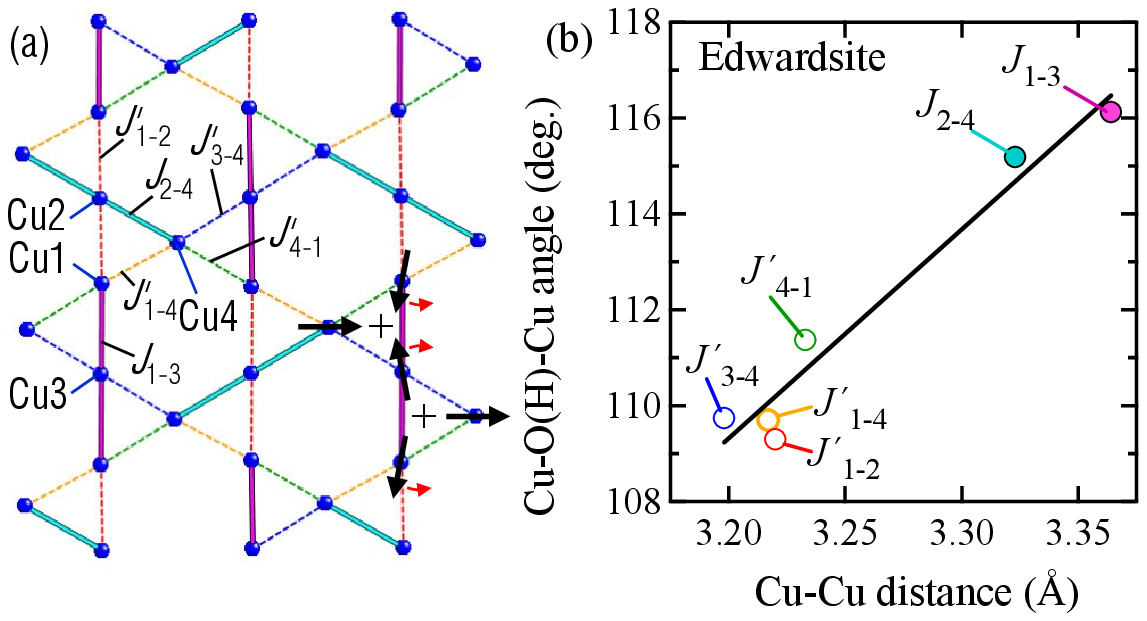}
\caption{(a) Magnetic interactions on the deformed kagome lattice in edwardsite. (b) Cu-O(H)-Cu angles versus Cu-Cu distances for the six nonequivalent Cu-Cu bonds in edwardsite~\cite{Elliott}. The solid line is a visual guide. $J_{2-4}$ and $J_{1-3}$ with larger Cu-O-Cu angles, which may give stronger antiferromagnetic interactions, are represented by thick solid lines in (a), while $J^{\prime}_{1-2}$, $J^{\prime}_{1-4}$, $J^{\prime}_{3-4}$, and $J^{\prime}_{4-1}$ with smaller angles and probably weaker interactions are shown by thin broken lines in (a). A possible \textbf{q} = 0 spin structure is also shown in the lower-right part of (a). }
\label{F6}
\end{center}
\end{figure}

DM interactions may be crucial for the occurrence of magnetic order in vesignieite as well as in edwardsite~\cite{Vesi4}. The DM vector $D$ takes two components: $D_{\perp}$ and $D_{\parallel}$, which are perpendicular and parallel to the kagome lattice, respectively. The former stabilizes the in-plane 120$^{\circ}$ order with a \textbf{q} = 0 propagation vector, while the latter gives rise to the canting of spins to the out-of-plane direction, resulting in a weak ferromagnetic moment~\cite{DM1}. Theory predicts that $D_{\perp}$/$J \sim$ 0.1 is the critical point between a spin liquid and a magnetic order~\cite{DM2}. $D$/$k_{\mathrm{B}} \sim$ 6 K in vesignieite suggests that it lies on the order side~\cite{Vesi4}. $D$ in edwardsite may be larger, because the $M_{\mathrm{WF}}$ of edwardsite is almost twice as large as that of vesignieite. Hence, it is probable that $D_{\perp}$ is larger than 0.1$J$ in edwardsite, which stabilizes the \textbf{q} = 0 order with spin canting. 

Finally, we discuss the effects of the deformation of the kagome lattice on the magnetic properties of edwardsite. As shown in Fig. 6(a), Cu$^{2+}$ ions occupy four different crystallographic sites on the kagome plane, giving six nonequivalent Cu-Cu bonds with different magnetic couplings. A superexchange interaction via a Cu-O(H)-Cu pathway must be crucial in each bond, because all four lobes of the $x^2 - y^2$ orbital at every Cu site point to the OH$^-$ ions, as shown in Fig. 1(b). This interaction must be sensitive to the Cu-O-Cu angle $\theta$: ferromagnetic for $\theta \lesssim$ 95$^{\circ}$ and antiferromagnetic for $\theta \gtrsim$ 95$^{\circ}$~\cite{Mizuno}. All the magnetic interactions in edwardsite must be antiferromagnetic, because the Cu-O(H)-Cu angles are large, i.e., between 109$^{\circ}$ and 116$^{\circ}$~\cite{Elliott}, as shown in Fig. 6(b). 

Note that there are two groups with small and large Cu-O-Cu angles in Fig. 6(b). This suggests that the six antiferromagnetic couplings are classified into two groups, i.e., two stronger ones with $J$ and four weaker ones with $J^{\prime}$. Then, the kagome lattice of edwardsite approximately consists of $J$-$J^{\prime}$-$J^{\prime}$ isosceles triangles that are arranged to form linear trimers with $J$. It is expected for classical spins in such a distorted $J$-$J^{\prime}$-$J^{\prime}$ triangle that two spins coupled by stronger $J$ will tend to cant from the 120$^{\circ}$ structure so as to align more antiparallel to each other, as shown in Fig. 6(a)~\cite{DistKagome}. This canting is not allowed in the $\sqrt{3} \times \sqrt{3}$ structure, but allowed in the \textbf{q} = 0 structure with uniform chirality. Therefore, it is reasonable that a \textbf{q} = 0 structure is realized in edwardsite. 

In summary, we show that edwardsite is a spin-1/2 deformed KAFM with an average $J$/$k_{\mathrm{B}}$ = 51.1 K and exhibits an antiferromagnetic order accompanied by a small ferromagnetic moment below $T_{\mathrm{N}}$ = 4.3 K. The magnetic structure is suggested to be a canted 120$^{\circ}$ order with a \textbf{q} = 0 propagation vector, which is preferred by a strong DM interaction and the deformation of the kagome lattice. Further experiments, particularly under high magnetic fields, would show interesting phenomena in this unique KAFM.


\end{document}